\documentstyle[11pt]{article}

\begin{document}

\small \hoffset = -1truecm \voffset = -2truecm

\title{\Large\bf Variable separation approach for a differential-difference
system: special Toda equation}

\author{\small  Xian-min Qian$^{1, 2, 4}$, Sen-yue Lou$^{1, 3}$,Xing-biao Hu$^{4}$\\
\small \it $^1$ Physics Department of Shanghai Jiao Tong University,Shanghai,200030,P.R.China\\
\small \it $^2$ Physics Department of Shaoxing college of arts and
sciences, Shaoxing, 312000, P. R. China\\\small \it $^3$ Physics
Department of Ningbo University,Ningbo,
315211, P.R.China\\
\small \it $^4$ State Key Laboratory of Scientific and Engineering
Computing,\\\small \it Institute of Computational Mathematics and
Scientific Engineering Computing, \\ \small \it Academy of
Mathematics and System Sciences,Academia Sinica, \\
\small \it PO Box 2719, Beijing 100080, People's Republic of
China}

\date{}

\maketitle

\begin{quotation}

\centerline{\bf Abstract} A multi-linear variable separation
approach is developed to solve a differential-difference Toda
equation. The semi-discrete form of the continuous universal
formula is found for a suitable potential of the
differential-difference Toda system. Abundant semi-discrete
localized coherent structures of the potential can be found by
appropriately selecting the arbitrary functions of the
semi-discrete form of the universal formula.

\leftline{\bf PACS numbers:\rm 02.30.Ik, 05.45.Yv}
\end{quotation}

\section{Introduction}

It is very difficult to find explicit exact solutions both for
nonlinear partial differential equations (PDEs) and for nonlinear
differential-difference equations (DDEs). In linear mathematical
physics, the Fourier transform and the variable separation
approach are two of the most effective ways to find exact
solutions of linear equations. In nonlinear mathematical physics,
the so-called inverse scattering transform (IST) has served as
``nonlinear" Fourier transform for nonlinear integrable models.
However, it is very hard to extend the variable separation
approach to nonlinear cases even for integrable systems.

Recently, several kinds of ``variable separation" approaches, say,
the classical method, the differential St\"ackel matrix approach
\cite{Miller}, the geometric method \cite{Do1}, the ansatz-based
method \cite{Rz1, Do1}, functional variable separation
approach\cite{Zh1}, the derivative dependent functional variable
separation approach\cite{Zsl}, the formal variable separation
approach (nonlinearization of the Lax pairs or symmetry
constraints) \cite{Lou3} and the multilinear variable separation
approach (MVSA) \cite{Lou1}--\cite{Lou4} are in progress. Among
these approaches, the MVSA may be the most powerful method to
solve many (2+1)-dimensional systems both for integrable
models\cite{Lou1}--\cite{Lou4} and for nonintegrable
systems\cite{TangLou}. In \cite{Lou4}, a quite universal formula
\begin{equation}\label{0}
U\equiv \frac{2(a_1a_2-a_0a_3) q_{y}p_{x}}{(a_0+a_1 p+a_2 q+a_3 p
q)^2},\
\end{equation}
is established to describe suitable physical quantities for
various (2+1)-dimensional models solvable via MVSA. In (\ref{0}),
$a_0,\ a_1,\ a_2$ and $a_3$ are arbitrary constants and $p$ is an
arbitrary function of $\{x,\ t\}$ for all of the known MVSA
solvable models, while $q$ of (\ref{0}) may be an arbitrary
function of $\{y,\ t\}$ for some of MVSA solvable models, or an
arbitrary solution of a Riccati equation for some other MVSA
solvable models. Because some arbitrary characteristics, lower
dimensional functions (like $p$), have been included in the
universal formula (\ref{0}), by selecting them appropriately,
abundant localized structures like the multiple solitoffs,
dromions, lumps, breathers, instantons, ring solitons and chaotic
and fractal patterns have been found\cite{Lou4}.

In this paper, we are interested in the following important
question: Can the MVSA be extended to solve some nonlinear DDEs?
In section 2, the MVSA is outlined for arbitrary DDEs. In sections
3 and 4, the MVSA is applied to a special differential-difference
Toda equation (SDDTE). A semi-discrete form of the universal
formula (\ref{0}) for a suitable potential of the SDDTE is given
in section 5. Starting from the semi-discrete form of the
universal formula, the abundant semi-discrete localized
excitations can be found for the potential of the SDDTE. Some
special semi-discrete localized excitations of the model are also
plotted in section 5. Section 6 contains a short summary and
discussions.

\section{Outline of the MVSA}

Consider a given DDE of the following type
\begin{eqnarray}\label{DDE}
&&F(x_i, n_j,u(x_i,n_j-k, ...),u_{x_i}(x_i,n_j-k, ...),...,i=1,\
...,N_1,j=1,...,N_2,k=0,\pm1,\
\pm2,...)\nonumber\\
&& \equiv F(u)=0,
\end{eqnarray}

where $x_i,\ i=1,2,...N_1$ are continuous variables and $n_j,\
j=1,2,...N_2$ are discrete variables.
As in the continuous case, the MVSA can be performed for DDEs via
the following standard procedures.\\
(i). Multilinearize the original DDE (\ref{DDE}) by using a
suitable B\"acklund transformation.
Usually, the resulting equations are the bilinear equations for integrable systems. \\
(ii). Choose a seed solution of the B\"acklund transformation
as general as possible with one or more arbitrary functions. \\
(iii). Make a suitable variable separation ansatz with some
variable separated functions. Usually, the ansatz is just the
generalization of the Hirota's two-soliton
solution.\\
(iv). Substitute the ansatz into the multilinear equations and
separate the resulting equations to several variable separated ones. \\
(v). Solve the variable separated equations. Usually, to solve the
variable separated PDEs and/or DDEs is still very difficult for
any fixed seed solution. However, one can treat the problem in an
alternative easy way: The variable separated functions appeared in
the ansatz can be considered as arbitrary functions and then the
function(s) appeared in the seed solution
should be fixed from the variable separated equations. \\
(vi). Find a suitable field or potential which possesses a special
variable separated solution described by the universal formula
(\ref{0}) for PDEs
($N_2=0$ in (\ref{DDE})) or a semi-discrete form of (\ref{0}) for DDEs. \\
(vii). Discuss the possible semi-discrete localized excitations by
selecting the arbitrary functions appropriately.

To see the details on the procedures of the MVSA for the DDE
systems, we take the SDDTE as a simple concrete example in the
remained sections.

\section{SDDTE and its generalized bilinear form}

In nonlinear discrete and semi-discrete physics systems, the most
famous and important systems are the so-called Toda systems which
are widely used in physics\cite{ZJ}. In this paper, we only
consider a special differential-difference Toda equation (SDDTE)
\begin{equation}\label{toda}
Q(n)_{yt}=\exp[Q(n+1)-Q(n)][Q(n+1)+Q(n)]_y-\exp[Q(n)-Q(n-1)][Q(n)+Q(n-1)]_y,
\end{equation}
where $Q(n)\equiv Q(n,\ y,\ t)$ is a function of the discrete
variable $n$ and the continuous variables $\{y,\ t\}$. The SDDTE
(\ref{toda}) was firstly derived by Cao, Geng and Wu in a
remarkable paper \cite{SDDTE}. Some interesting integrable
properties of the SDDTE (\ref{toda}) have been given in
\cite{SDDTE} and \cite{Hu}.

To solve the SDDTE (\ref{toda}) via MVSA, one can use the
following dependent variable transformation
\begin{equation}\label{BT}
Q(n)=\rho(n)+\ln\left(\frac{f(n+1)}{f(n)}\right)
\end{equation}
to bilinearize it. In the transformation (\ref{BT}),
$\rho(n)\equiv \rho(n,t)$, the seed solution of the SDDTE has been
selected as an arbitrary function of $\{n,\ t\}$ for convenience
later.

Substitution of the dependent variable transformation (\ref{BT})
into the SDDTE yields
\begin{eqnarray}\label{fbi}
& &(T_+-1)\left(\frac{D_yD_tf(n)\cdot
f(n)-2\exp[\rho(n)-\rho(n-1)]D_y \exp(D_n)f(n)\cdot
f(n)}{2f(n)^2}\right)=0,
\end{eqnarray}
where $T_+$ is a shift operator, i.e.,$T_+ F(n)=F(n+1)$, and the
Hirota's bilinear differential operator $D_y^mD_t^k$ and the
bilinear difference operator $\exp{D_n}$  are defined by
$$\left. D_y^mD_t^k a \cdot b\equiv \left(\frac{\partial}{\partial{y}}
-\frac{\partial}{\partial {y'}}\right)^m
\left(\frac{\partial}{\partial{t}}-\frac{\partial}{\partial
{t'}}\right)^ka(y,t)b(y',t')\right|_{y=y',t=t'},$$
$$\exp(\delta D_n)a(n)\cdot
b(n)\equiv \left.
\exp\left[\delta\left(\frac{\partial}{\partial{n}}-\frac{\partial}{\partial
{n'}}\right)\right]a(n)b(n)\right|_{n=n'}\equiv
a(n+\delta)b(n-\delta).$$

Multiplying (\ref{fbi}) by the inverse operator of $T_+-1$ leads
to a generalized bilinear SDDTE
\begin{eqnarray}\label{mfbi}
D_yD_tf(n)\cdot f(n)-J(n,t)D_y \exp(D_n)f(n)\cdot
f(n)+R(y,t)f(n)^2=0,
\end{eqnarray}
 where
 \begin{eqnarray}\label{rho}
J(n)\equiv 2\exp[\rho(n)-\rho(n-1)]
\end{eqnarray}
 is an arbitrary function of $\{n,\ t\}$ and $R(y,t)$, the kernel of the difference operator $T_+-1$,
 is an arbitrary function of $\{y,\ t\}$.

\section{Variable separation solution of the SDDTE}

\qquad  In order to find some exact solution of (\ref{mfbi}),
similar to the continuous cases\cite{Lou1}--\cite{Lou4}, we look
for the solutions of (\ref{mfbi}) in the form
\begin{equation}\label{fpq}
f(n)=a_0+a_1 p(n,t) +a_2 q(y,t)+a_3 p(n,t) q(y,t),
\end{equation}
where $ a_0,\ a_1,\ a_2$ and $a_3$ are arbitrary constants and the
variable separation functions $q(y,t)\equiv q$ and $p(n,t)\equiv
p(n)$ are only functions of $\{y,\ t\}$ and $\{n,\ t\}$
respectively. (\ref{fpq}) looks like the Hirota's two soliton form
when $q$ and $p(n)$ are exponential functions. Substituting the
ansatz (\ref{fpq}) into (\ref{mfbi}), we have
\begin{eqnarray}\label{todapq}
&&-2(a_2+a_3p(n))^2q_yq_t+2(a_3a_0-a_2a_1)q_yp_{nt}+2(a_2+a_3p(n))
(a_0+a_1p(n)+a_2q+a_3p(n))q_{yt}{\nonumber}\\
&&-J(n)(p(n+1)-p(n-1))(a_3a_0-a_2a_1)q_y+R(a_0+a_2q+a_1p(n)+a_3qp(n))^2=0
\end{eqnarray}
Because $p(n)$ and $J(n)$ are only functions of $\{n,\ t\}$ and
$q$ and $R$ are only functions of $\{y,\ t\}$, (\ref{todapq}) can
be separated into the following two equations,
\begin{eqnarray}
&&q_t=(a_0+a_2q)^2c_1+(a_1+a_3q)^2c_2+(a_0+a_2q)(a_1+a_3q)c_3,\label{todaqt}\\
&&p_t(n)=(a_0a_3-a_1a-2)(c_2-c_3p(n)+c_1p(n)^2)+\frac12J(n)(p(n+1)-p(n-1)),\label{todapt}
\end{eqnarray}
when the arbitrary function $R$ is selected as
\begin{eqnarray}\label{R}
R=-2(c_1a_2^2+c_2a_3^2+c_3a_2a_3)q_y,
\end{eqnarray}
with $c_1\equiv c_1(t),\ c_2\equiv c_2(t)$ and $c_3\equiv c_3(t)$
being arbitrary functions of $t$.

In principle, as long as the arbitrary functions $c_1,\ c_2,\ c_3$
and $\rho(n)$ (and then $J(n)$) are fixed, we can obtain the
corresponding special solutions of the (\ref{todaqt}) and
(\ref{todapt}) and then the special solutions of the SDDTE
(\ref{toda}). However, it is still very difficult to solve the
nonlinear DDE (\ref{todapt}) for fixed nonzero $J(n)$.
Fortunately, as in the continuous cases discussed in
\cite{Lou1}--\cite{Lou4}, because of the arbitrariness of the
function $J(n)$, we can treat the problem alternatively. We can
consider the function $p(n)$ as an arbitrary function of the
variables $n$ and $t$ and fix the function $J(n)$ from the
equation (\ref{todapt}). The result reads
\begin{eqnarray}\label{Jn1}
J(n)=\frac{2}{p(n-1)-p(n+1)}[(a_0a_3-a_1a-2)(c_2-c_3p(n)+c_1p(n)^2)-p_t(n)].
\end{eqnarray}

It should be pointed out that the Riccati equation (\ref{todaqt})
is totally same as that of the asymmetric Nizhnik-Novikov-Veselov
(ANNV) equation\cite{Lou4}. To find out some special solutions of
(\ref{todaqt}), one may select the arbitrary functions
appropriately.
Here we list two special selections. \\
(1). If we write $c_1,\ c_2$ and $c_3$ as
\begin{eqnarray}
c_1&=&\frac{a_3^2A_{2t}}{(a_1a_2-a_0a_3)^2}-\frac{a_3(a_1+a_3A_2)A_{1t}}{(a_1a_2-a_0a_3)^2A_1}
-\frac{(a_1+a_3A_2)^2A_{3t}}{(a_1a_2-a_0a_3)^2A_1},\label{c11}\\
c_2&=&\frac{a_2^2A_{2t}}{(a_1a_2-a_0a_3)^2}-\frac{a_2(a_0+a_2A_2)A_{1t}}{(a_1a_2-a_0a_3)^2A_1}
-\frac{(a_0+a_2A_2)^2A_{3t}}{(a_1a_2-a_0a_3)^2A_1},\label{c21}\\
c_3&=&\frac{(a_0a_3+a_1a_2+2a_2a_3A_2)A_{1t}}{(a_1a_2-a_0a_3)^2A_1}-\frac{2a_2a_3A_{2t}}
{(a_1a_2-a_0a_3)^2}\nonumber \\
&&+2\frac{(a_0+a_2A_2)(a_1+a_3A_2)A_{3t}}{(a_1a_2-A)^2A_1}\label{c31}
\end{eqnarray}
with $A_1\ \equiv\ A_1(t), A_2\ \equiv\ A_2(t)$ and $A_3\ \equiv\
A_3(t)$ being arbitrary functions of $t$, then the general
solution of (\ref{todaqt}) with (\ref{c11})-(\ref{c31}) reads
\begin{eqnarray}\label{q1}
q=\frac{A_1}{A_3+F_1(y)}+A_2.
\end{eqnarray}
where $F_1\ \equiv\ F_1(y)$ is an arbitrary function of $y$.\\
(2). If we select $c_1,\ c_2$ and $c_3$ as
\begin{eqnarray}
c_1&=&\frac{a_3^2b_{0t}}{(a_1a_2-a_0a_3)^2}-\frac{a_3(a_1+a_3b_0)b_{1t}}{(a_1a_2-a_0a_3)^2b_1}
-\frac{[(a_1+a_3b_0)^2-b_1^2a_3^2]b_{2t}}{(a_1a_2-a_0a_3)^2b_1},\label{c12}\\
c_2&=&\frac{a_2^2b_{0t}}{(a_1a_2-a_0a_3)^2}-\frac{a_2(a_0+a_2b_0)b_{1t}}{(a_1a_2-a_0a_3)^2b_1}
-\frac{[(a_0+a_2b_0)^2-a_2^2b_1^2]b_{2t}}{(a_1a_2-a_0a_3)^2b_1},\label{c22}\\
c_3&=&\frac{(a_0a_3+a_1a_2+2a_2a_3b_0)b_{1t}}{(a_1a_2-a_0a_3)^2b_1}-\frac{2a_2a_3b_{0t}}
{(a_1a_2-a_0a_3)^2}\nonumber\\
&&+2\frac{[(a_0+a_2b_0)(a_1+a_3b_0)-a_2a_3b_1^2]b_{2t}}{(a_1a_2-A)^2b_1}\label{c32}
\end{eqnarray}
with $b_0\ \equiv\ b_0(t),\ b_1\ \equiv\ b_1(t)$ and $b_2\ \equiv
b_2(t)$ being arbitrary functions of $t$, then the general
solution of (\ref{todaqt}) with (\ref{c12})-(\ref{c32}) reads
\begin{eqnarray}\label{q2}
q\ =\ {b_1}\tanh{(b_2+F_2(y))}+b_0
\end{eqnarray}
with $F_2\ \equiv\ F_2(y)$ being an arbitrary function of $y$.

\section{Abundant coherent structures for a potential of the SDDTE}

Substituting all the results of the last section into (\ref{BT}),
one can get many kinds of exact solutions for the field $Q$ of the
SDDTE. In continuous cases, it has been pointed out that for every
of the MVSA solvable systems listed in \cite{Lou4},
there exits a quantity that can be described by the universal formula
(\ref{0}). Now an important question is: \\
\bf Is there a suitable potential for the SDDTE that can be
described by a suitable semi-discrete form of the universal
formula (\ref{0})? \rm

Fortunately, it is straightforward to prove that if we define a
potential of the SDDTE as
\begin{eqnarray}\label{u}
u\equiv -2Q_y(n),
\end{eqnarray}
then
\begin{eqnarray}\label{Un}
u=U(n)\equiv \frac{2q_y(a_2a_1-a_3a_0)(p(n+1)-p(n))}
{(a_0+a_2q+a_1p(n)+a_3qp(n))(a_0+a_2q+a_1p(n+1)+a_3qp(n+1))}.
\end{eqnarray}
It is clear that the function $U(n)$ defined in (\ref{Un}) is just
one suitable semi-discrete form of the continuous universal
quantity $U$ given in (\ref{0}). We say the function $U(n)$ is
semi-discrete means that it is discrete in one direction and
continuous in other direction.

Now starting from the semi-discrete form of the universal
quantity, we can obtain abundant semi-discrete localized
excitations for the SDDTE by selecting the arbitrary functions
appropriately.

The detailed studies show us that the semi-discrete localized
structures for the potential $u$ expressed by the semi-discrete
form of the universal quantity are very similar to the continuous
ones which have been discussed in Ref. \cite{Lou4}. So in this
paper we will not discuss all the possible localized excitations
but only list some particular examples.

\leftline{\bf Example 1. Resonant semi-discrete dromion and
solitoff solutions.}

If we restrict the functions $p(n)$ and $q$ of (\ref{0}) as
\begin{eqnarray}\label{offp}
&&p(n)=\sum_{i=1}^N\exp(k_in+\omega_it+x_{0i})\ \equiv\
\sum_{i=1}^N\exp(\xi_i)\\
&&q=\sum_{i=1}^M\exp(K_{i}y+y_{0i})\sum_{j=1}^J\exp(\Omega_jt+t_{0j}),
\label{offq}
\end{eqnarray}
where $x_{0i},\ y_{0i},\ t_{0j},\ k_i,\ \omega_i,\ K_i$ and
$\Omega_i$ are arbitrary constants and $M,\ N$ and $J$ are
arbitrary positive integers, then we have a single resonant
semi-discrete dromion solution or semi-discrete multiple solitoff
solutions. The selection (\ref{offq}) is related to the selections
on the functions of $A_i,\ i=1,\ 2,\ 3, \ F_1$ of (\ref{q1}) as
\begin{eqnarray}
&&A_3=A_2=0, \label{A231}\\
&&A_1=\sum_{j=1}^J\exp(\Omega_jt+t_{0j}),\label{A11}\\
&&F_1=\frac1{\sum_{i=1}^M\exp(K_{i}y+y_{0i})},\label{F11}
\end{eqnarray}
and the $c_i,\ i=1,\ 2,\ 3$ are given by (\ref{c11})--(\ref{c31})
with (\ref{A231}) and (\ref{A11}).

In Fig.1, we plot four typical structures caused by the resonant
effects of four straight-line semi-discrete soliton solutions.

Fig. 1a shows the structure of a first type of single resonant
semi-discrete dromion solution shown by (\ref{u}) with
(\ref{offp}), (\ref{offq}),
\begin{equation}\label{sol11}
M=N=2, J=k_1=K_1=1,\ k_2=K_2=\frac13,\ a_0=1,\ a_1=a_2=10,\
a_3=\frac12
\end{equation}
and
\begin{equation}\label{xy0}
x_{01}\ =\ y_{01}\ =t_{01}\ =\ x_{02}\ =y_{02}\ =\ 0
\end{equation}
at $t=0$.

Fig. 1b is a plot of a single resonant semi-discrete solitoff
solution shown by (\ref{u}) with (\ref{offp}), (\ref{offq}),
(\ref{xy0}) and
\begin{equation}\label{sol13}
M=N=2, J=k_1=K_1=1,\ k_2=K_2=\frac13,\ a_0=1,\ a_1=a_2=3, \
a_3=0,\ t=0.
\end{equation}

Fig. 1c shows the structure of a second type of single
semi-discrete resonant dromion solution shown by (\ref{u}) with
(\ref{offp}), (\ref{offq}), (\ref{xy0}) and
\begin{equation}\label{sol12} M=N=2, J=k_1=-K_1=1,\
-k_2=K_2=\frac13,\ a_0=a_2=10,\ a_3=1, a_1=\frac12, \ t=0.
\end{equation}

Fig. 1d is a plot of a four solitoff solution shown by (\ref{u})
with (\ref{offp}), (\ref{offq}), (\ref{xy0}) and
\begin{equation}\label{sol22}
M=N=2, J=k_1=-K_1=1,\ -k_2=K_2=\frac13,\ a_0=1, \ a_1=a_2=3,\
a_3=0, t=0.
\end{equation}

\input epsf
\begin{figure}
\centering \epsfxsize=7cm\epsfysize=5cm \epsfbox{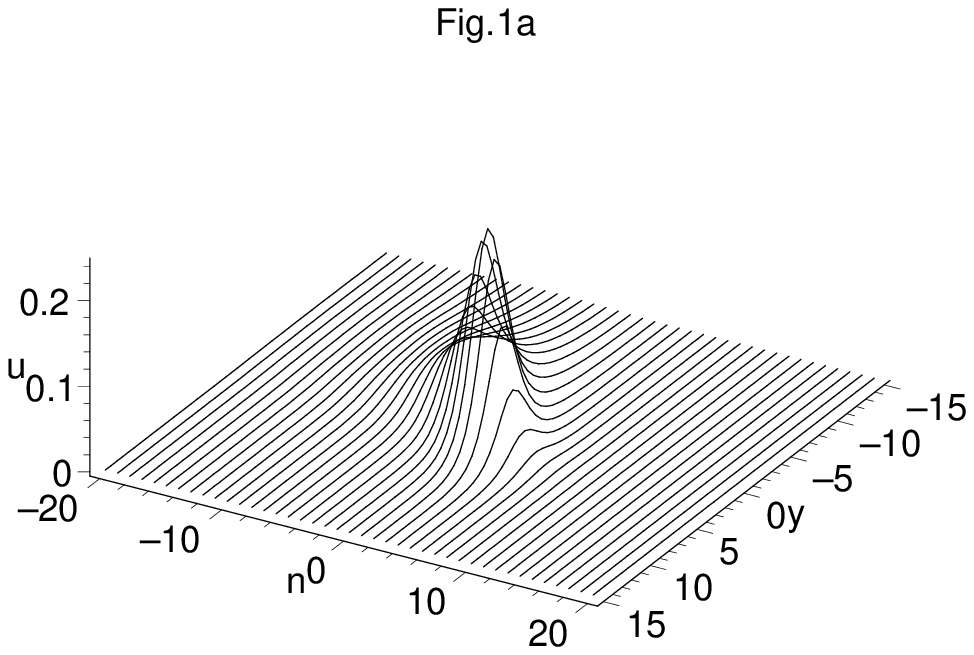}
\centering \epsfxsize=7cm\epsfysize=5cm \epsfbox{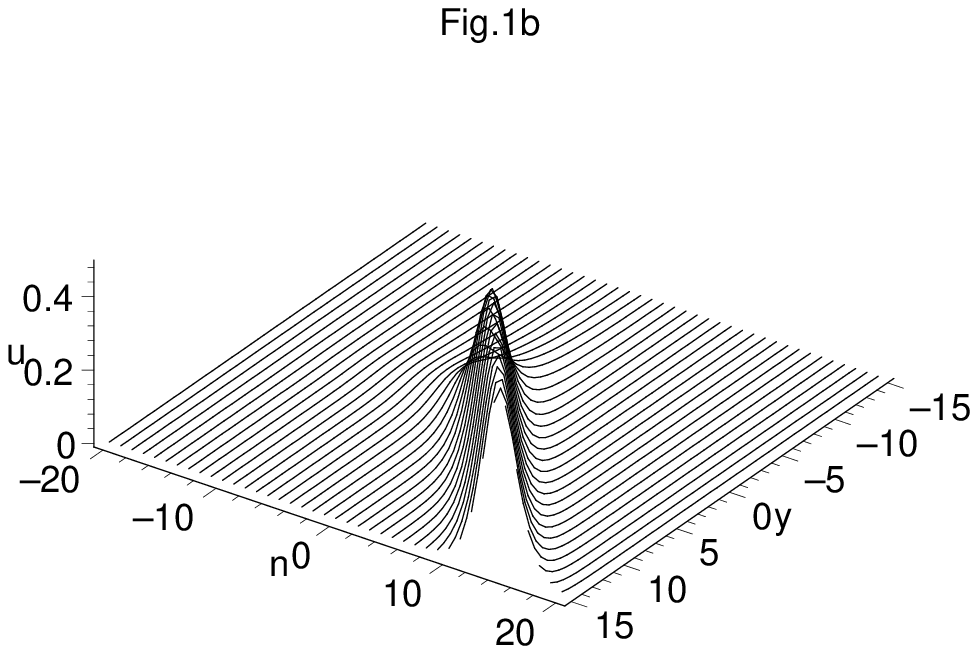}
\centering \epsfxsize=7cm\epsfysize=5cm \epsfbox{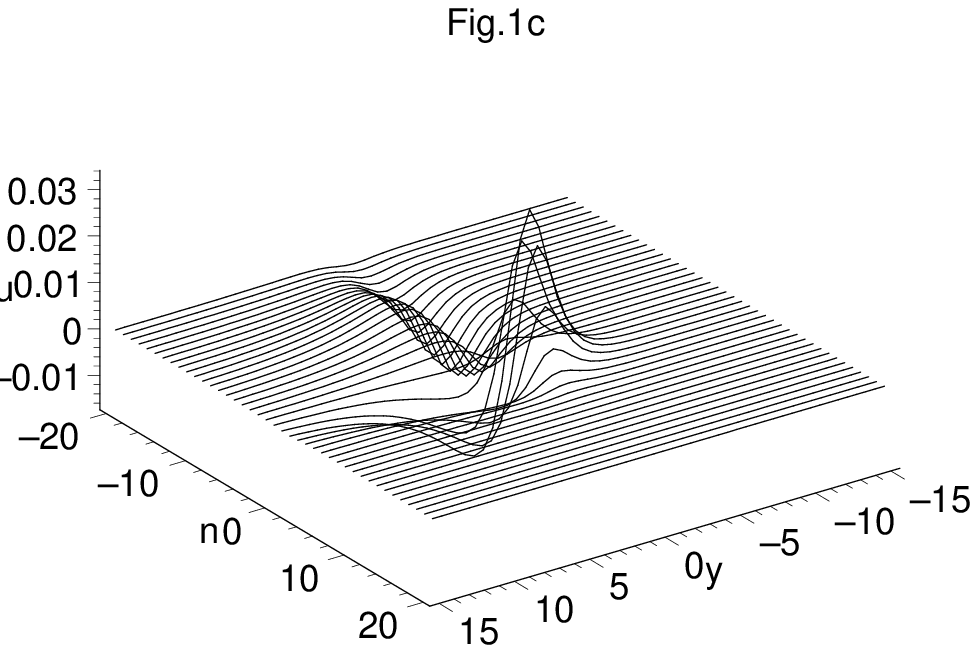}
\centering \epsfxsize=7cm\epsfysize=5cm \epsfbox{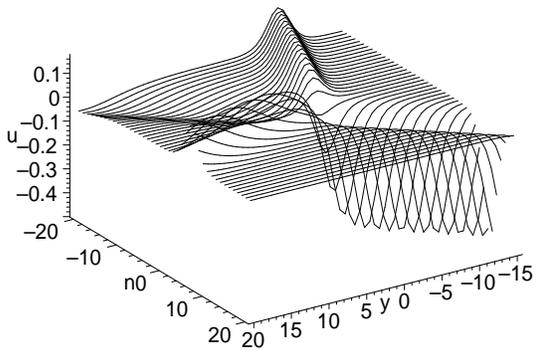}
\caption{\footnotesize Four typical semi-discrete structures of
SDDTE for the potential $u$ expressed by (\ref{u}) with
(\ref{offp}) and (\ref{offq}). (a). A special single-peak
semi-discrete resonant dromion solution. (b). A single
semi-discrete solitoff solution. (d). A multi-peak semi-discrete
dromion solution. (b). A semi-discrete four-solitoff solution. }
\end{figure}

\leftline{\bf Example 2. Semi-discrete oscillating dromions and
lumps.}

If some periodic functions in space variables are included in the
functions $p(n)$ and $q$ of (\ref{u}), we may obtain some types of
semi-discrete multi-dromion and multi-lump solutions with
oscillating tails. The oscillated lump solution plotted in Fig. 2
is related to
\begin{eqnarray}\label{tail}
\displaystyle q=\frac1{1+[y(\cos(y)+5/4)]^2},\
p(n)=\frac1{1+(n-ct)^2},
\end{eqnarray}
\begin{eqnarray}\label{tail1}
a_0=a_3=1,\ a_1=a_2=5
\end{eqnarray}
at $t=0$.

\input epsf
\begin{figure}
\centering \epsfxsize=7cm\epsfysize=5cm \epsfbox{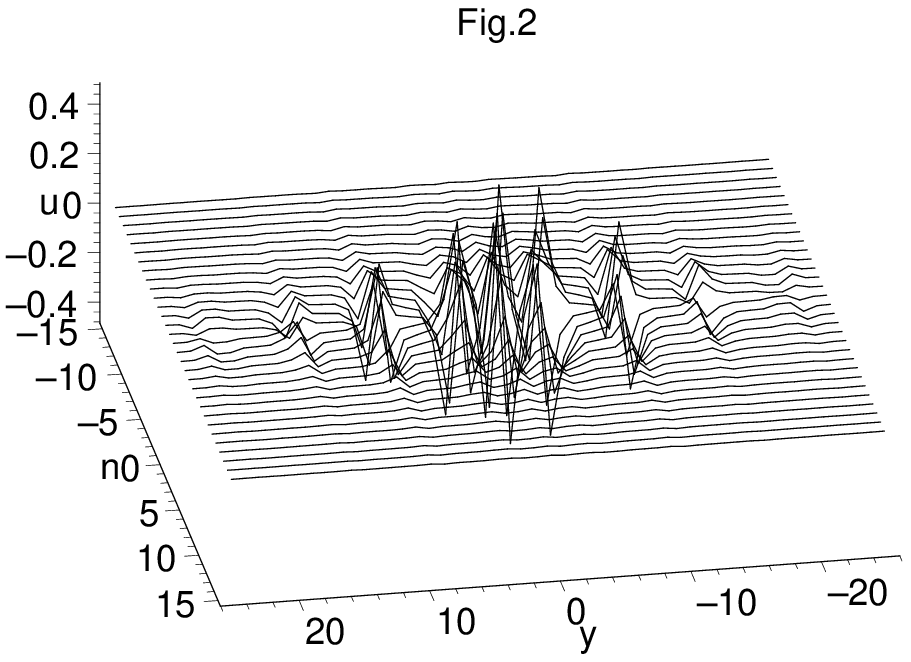}
\caption{\footnotesize Plot of a special oscillating lump solution
of the SDDTE for the potential $u$ expressed by (\ref{u}) with
(\ref{tail}) and (\ref{tail1}) at $t=0$.}
\end{figure}

\leftline{\bf Example 3. Multiple ring soliton solutions.}

In high dimensions, in addition to the point-like localized
coherent excitations, there may be some other types of physically
significant localized excitation. Recently, we have found some
different kinds of ring soliton solutions which are not
identically equal to zero at some closed (2+1)-dimensional and
(3+1)-dimensional curves and decay exponentially away from the
curves\cite{2dsg, Lou4, LouYu}.

In Fig. 3, a typical saddle type semi-discrete ring soliton
solution is plotted for the potential $u$ with the selections
\begin{equation}\label{ring}
q=\exp\left(-\frac{y^2}{80}+5\right),\
p(n)=\exp\left(\frac{(n-ct)^2}{80}\right),
\end{equation}
and
\begin{equation}\label{ringc}
a_0=a_3=0,\ a_1=a_2=5,
\end{equation}
at $t=0$.

\input epsf
\begin{figure}
\centering \epsfxsize=7cm\epsfysize=5cm \epsfbox{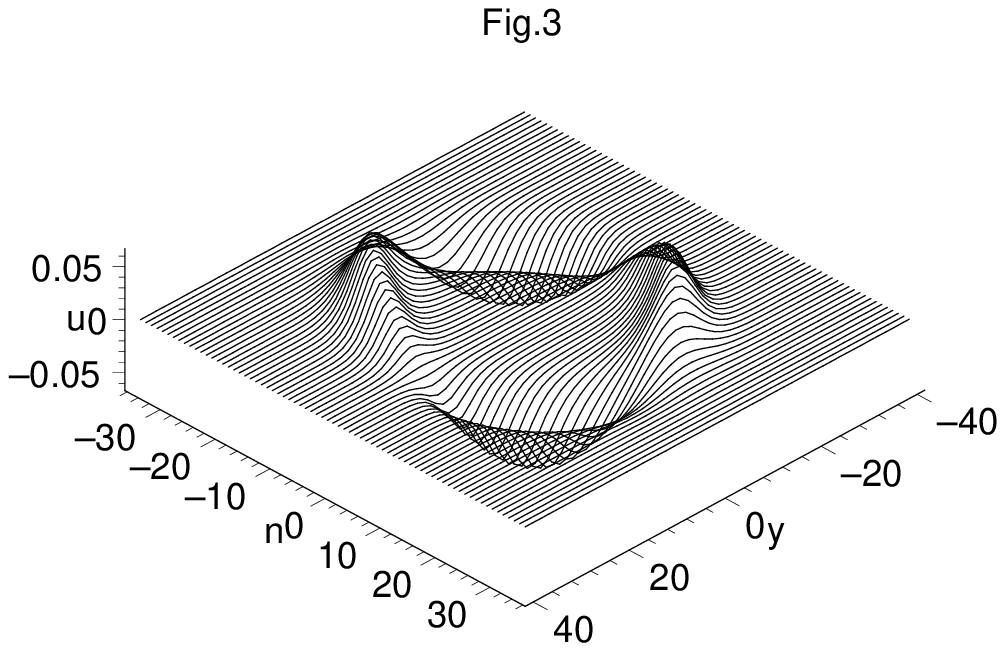}
\caption{\footnotesize Plot of a typical single saddle type
semi-discrete ring soliton solution for the potential $u$ of the
SDDTE with the selections (\ref{ring}) and (\ref{ringc}) at
$t=0$.}
\end{figure}

Because of the existence of the arbitrariness in the expression
(\ref{u}) for the potential $u$, there exist many kinds of
multiple semi-discrete ring soliton solutions. Since the
situations are quite similar to those in the continuous cases, we
do not discuss them further. However it is worth to mention that
as pointed out in Refs. \cite{Lou4} and \cite{TangLou}, the
interactions among the travelling ring soliton solutions are
completely elastic without phase shift.

\section{Summary and discussion}
\qquad In the previous studies\cite{Lou1}--\cite{Lou4}, we have
successfully solved several famous (2+1)-dimensional nonlinear
continuous integrable models via a multilinear variable separation
approach (MVSA). In this paper, we have further extended the MVSA
to solve nonlinear differential-difference systems. Taking a
special differential-difference Toda equation (SDDTA) as a
concrete example, we have successfully solved the model via the
MVSA.

In continuous cases, a quite universal formula has been found to
describe suitable fields or potentials of MVSA solvable models.
For the SDDTA, a semi-discrete form of the universal formula is
found for a suitable potential. An arbitrary function related to
the discrete space variable is included in the semi-discrete form
of the universal formula. Another function included in the formula
is an arbitrary solution of the Riccati equation and the Riccati
equation is totally same as that in some continuous MVSA solvable
models.

By selecting the arbitrary functions appropriately, one can find
abundant semi-discrete localized excitations like the multiple
solitoffs, dromions, lumps, breathers, instantons and ring soliton
solutions. The semi-discrete localized solutions are quite similar
to those of continuous cases shown in \cite{Lou4}.

Though the MVSA have been successfully applied to the SDDTE, there
are various interesting and important problems are worth to
studying further. For instance, in continuous case, more than ten
models have been solved by the MVSA (see Ref. \cite{Lou4} and the
references therein). How many DDEs may be MVSA solvable? It is
also known that for a continuous integrable model, there exist
some different types of integrable discrete forms. Is the
semi-discrete form of the universal formula unique? In other
words, how universal the semi-discrete quantity (\ref{Un}) is? In
addition to the DDEs, there may be various fully discrete systems
in real physics. Can we extend the MVSA to solve some fully
discrete nonlinear systems?

\vskip.1in The authors are in debt to the helpful discussions with
Drs. X-y Tang, C-l Chen and S-l Zhang. The work is supported by
the Outstanding Youth Foundation (No.19925522), the Research Fund
for the Doctoral Program of Higher Education of China (Grant. No.
2000024832) and the Natural Science Foundation of Zhejiang
Province of China.

\end{document}